\documentclass[%
 aps,
 prl,%
 amsmath,
 amssymb,
reprint,%
groupedaddress,
]{revtex4-1}

\usepackage{color}%Color text
\usepackage[version=3]{mhchem}%Chemical Formulas
\usepackage{graphicx}% Include figure files
\usepackage{dcolumn}% Align table columns on decimal point
\usepackage{bm}% bold math
\usepackage{soul} %strickout 
\usepackage{hyperref}% hyperlink references
\hypersetup{
 linktocpage=true,
 pdfborderstyle={/S/S/W 1},
 hyperindex=true,
 bookmarks=true,
 bookmarksopen=true,
 bookmarksnumbered=true,
}% setup hyperlink

\begin{document}

\preprint{Yu Hao et alii}

\title{\textcolor{blue}{ Development of a broadband reflective T-filter for voltage biasing high-Q superconducting microwave cavities}}

%\thanks{Footnote to title of article.}

\author{Yu Hao}
\affiliation{Department of Physics, Syracuse University, Syracuse, New York 13244-1130, USA}

\author{Francisco Rouxinol}
\affiliation{Department of Physics, Syracuse University, Syracuse, New York 13244-1130, USA}

\author{M. D. LaHaye}
\email[]{Electronic mail: mlahaye@syr.edu}
\affiliation{Department of Physics, Syracuse University, Syracuse, New York 13244-1130, USA}

\date{\today}% It is always \today, today,
             %  but any date may be explicitly specified

\begin{abstract}
We present the design of a reflective stop-band filter based on quasi-lumped elements that can be utilized to introduce large dc and low-frequency voltage biases into a low-loss superconducting coplanar waveguide (CPW) cavity. Transmission measurements of the filter  are seen to be in good agreement with simulations and demonstrate insertion losses greater than $20\,{\rm dB}$ in the range of {\rm 3\,to\,10\,GHz}.  Moreover, transmission measurements of the CPW's fundamental mode demonstrate that loaded quality factors exceeding $10^5$ can be achieved with this design for dc voltages as large as ${\rm 20\,V}$ and for the cavity operated in the single-photon regime. This makes the design suitable for use in a number of applications including qubit-coupled mechanical systems and circuit QED.  
\end{abstract}

%\pacs{Valid PACS appear here}% PACS, the Physics and Astronomy
                             % Classification Scheme.
%\keywords{Hybrid quantum systems, circuit QED}%Use showkeys class option if keyword
                              %display desired
\maketitle

The integration of bias circuitry into high-quality (low-loss) microwave cavities for controlling embedded components is an important technical issue for a range of topics that includes research with qubit- and cavity-coupled mechanical systems,\cite{Irish2003,Blencowe2008,Clerk2007,Tian2008,LaHaye2009,Hertzberg2010,OConnell2010,Pirkkalainen2013,Xiang2013,Palomaki2013} circuit QED,\cite{Devoret2013f,Petersson2012} and quantum dynamics of nonlinear systems.\cite{Armour2013,Chen2014}  In these scenarios, the applied potentials and currents serve a variety of functions such as maintaining a device's operating state or establishing tunable electrostatic interactions between devices. However, if not carefully engineered, the introduction of the requisite circuitry can degrade the quality of a cavity through increased external circuit loading and radiative losses,\cite{Graaf2014} which generally have a detrimental effect on the given application.  

\begin{figure}

\includegraphics[ width=.7\columnwidth,keepaspectratio]{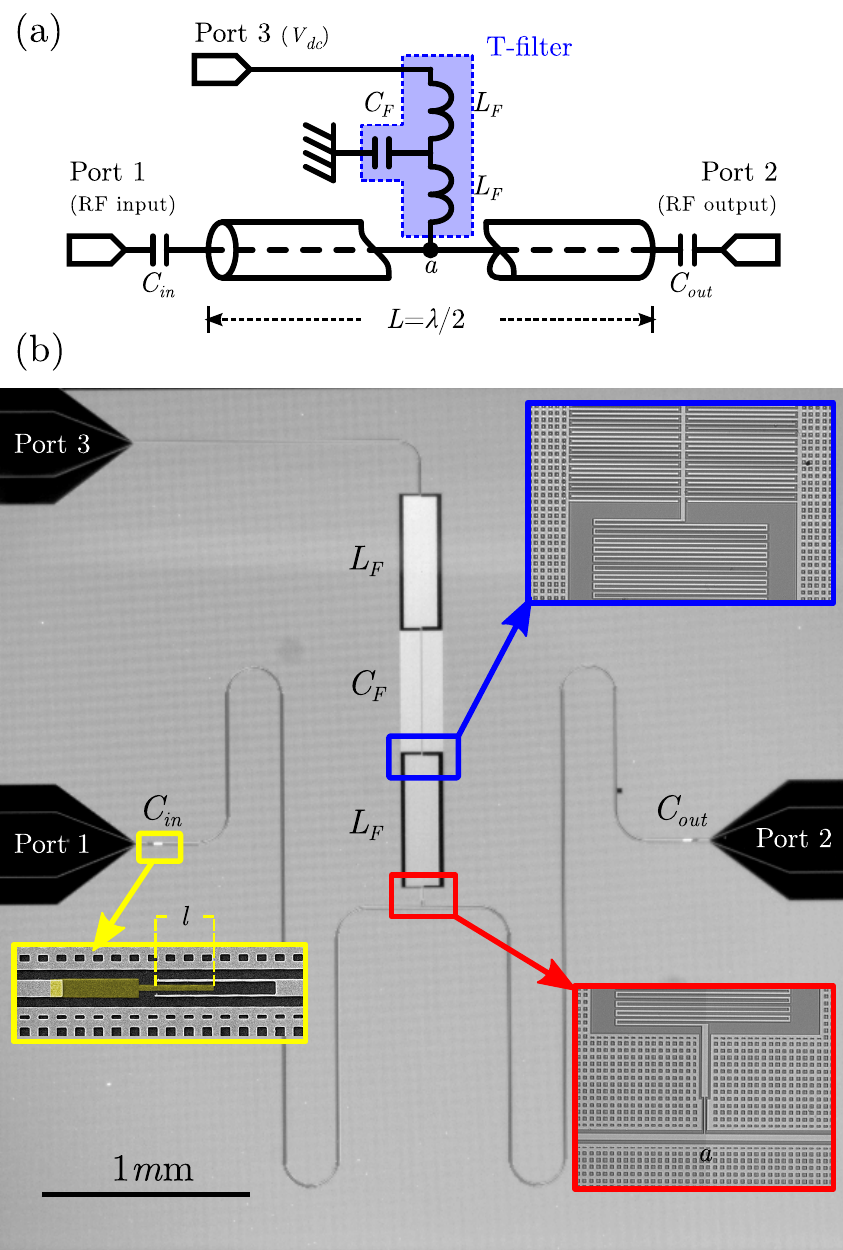}
\caption{ \footnotesize {\bf a)} Schematic of a CPW cavity and the reflective T-filter used for applying dc potentials to the CPW's center trace.  Ports 1 and 2 are the traditional input and output ports of the CPW cavity. Port 3 is the filtered dc port. {\bf b)} Optical image of one T-filtered CPW resonator studied here.  The device is representative of all the samples studied in this work with a 5.3 GHz fundamental CPW mode and symmetric coupling capacitors, $C_{in}$ and $C_{out}$ (inset, lower left), each composed of a single set of overlapping fingers.  The inset in the upper-right displays a close-up of the filter's interdigitated capacitor and one meander line inductor. The inset in the lower-right displays the connection between the filter and the CPW.}
    \label{fig1}
\end{figure}

In recent years, several solutions to this problem have been developed for dc biasing of superconducting coplanar waveguide (CPW) cavities. These cavities play an important role in the applications mentioned above due to their large electric field density\cite{Wallraff2004,Goppl2008} and high intrinsic quality factors ($\simeq 10^6$).\cite{Megrant2012} The solutions put forth for biasing such cavities utilize either half-wavelength or quarter-wavelength traces for band-stop filtering or high-impedance isolation of the bias circuitry's connections to the cavities.\cite{Chen2011,Li2013} This has enabled operation of biased CPWs with loaded quality factors in the range of $10^3$-$10^4$ for several applications.\cite{Petersson2012,Chen2014,Chen2011,Pozar2012} However, the reliance on wavelength-specific isolation geometries results in a narrow band of controlled isolation around a specific operating frequency, which could be problematic in instances where multiple devices with different or tunable energy scales are integrated into the cavity and thus require broadband isolation. Important examples of this include proposed techniques for the dispersive read-out of a qubit-coupled nanoresonator embedded inside a CPW cavity.\cite{Blencowe2008,Clerk2007} For such techniques to be feasible, the qubit and CPW, which can be detuned in energy by several ${\rm GHz}$, should have linewidths comparable to or smaller than the nanoresonator-qubit coupling strength, which is controlled by a large dc voltage (${\rm\sim10's\,V}$); this requires both the CPW and qubit to be well-isolated at their respective energies from the bias circuitry.    

Here we report a design that overcomes the challenge of obtaining broadband isolation in low frequency lines of a CPW cavity by utilizing a superconducting reflective T-filter.\cite{Pozar2012} This design provides high cavity isolation covering most of the frequency range typically used by elements in planar circuit QED architectures. Moreover, it allows for application of dc voltages as high as 20 V to the CPW's center trace\cite{voltage} with indiscernible distortion of the cavity's fundamental mode and no measurable increase in the mode's loss for quality factors exceeding $10^5$. This makes it an important new tool for investigations of quantum hybrid systems. 

A schematic of the filtered CPW cavity is shown in Fig. 1(a).  The cavity is formed in the standard manner\cite{Goppl2008} from a section of $Z_0=50\,\Omega$ planar transmission line that is isolated at its RF input and output ports (Ports 1 and 2 respectively) through the small capacitors $C_{in}$ and $C_{out}$. These capacitors set the ends of the cavity to be voltage anti-nodes, resulting in standing-wave resonances with wavelengths $\lambda= 2L/n$, where $L$ is the cavity length and $n$ is an integer.  In this work, we investigate the effect of the filter and low frequency circuitry on only the lowest-order mode, $\lambda=2L$, which has a voltage node at the mid-point of the cavity. 

In addition to the RF ports, a third port (Port 3) is connected to the mid-point of the cavity through a planar, low-pass T-filter, which allows for dc and low-frequency ($< 2\,{\rm GHz}$) voltage biasing of the cavity's center trace. The T-filter design, Figs. 1(a) and (b), consists of a single section of two meander-line inductors ($L_F$) located on opposite ends of an interdigitated shunt capacitor ($C_F$) to the ground plane.  Looking into the filter from port 3 (or from point a) it is intuitive how the filter element allows for low-frequency biasing and wide-band, high-frequency isolation of the cavity: At low frequencies ($\omega\ll\omega_F\equiv\sqrt{2/L_FC_F}$), the filter's inductive reactances are negligible compared to the capacitive reactance and load impedance, so any signal incident on the filter passes through unattenuated; on the other hand, when $\omega \gtrsim \omega_F$, the input impedance of the filter tends toward $ Z_{in}=j\omega L_F$, and hence signals incident on Port 3 are strongly reflected from the almost strictly imaginary impedance, i.e. $\gamma\equiv\frac{Z_{in}-Z_0}{ Z_{in}+Z_0}\approx1$. An analytical model using lumped-elements for the filter components predicts that the filter transmission ($S_{21}$) should roll off as $S_{21} \approx \frac{Z_0}{2 Z_F } \left(\frac{ \omega_F}{\omega}\right)^3$ for $ \omega>\omega_F$, where $Z_F\equiv \sqrt{(2L_F/C_F )}$ is the characteristic impedance of the filter. This yields an insertion loss that increases at 60 dB/decade. Of course, this estimate neglects the parasitic reactance of the filter elements, which plays a dominant role in limiting $IL$ over the frequency range explored here and is discussed in more detail below.

The CPW resonator, Fig. 1(b), is composed of a center line of width $6\,\mu$m and length of $L=11\,m$m, and it is separated from the ground planes by a gap of width $3\,\mu$m, resulting in a characteristic impedance of $50\,\Omega$ and resonance frequency $\omega_c/2\pi\approx5.3\,{\rm GHz}$. The interdigitated capacitor and the meander inductors that compose the filter have a central trace width and line-spacing of $2\,\mu$m. The capacitors and inductors were engineered to have nominal values of $L_F=10\,n$H  and $C_F=1.5\,p$F, corresponding to $\omega_F/2\pi\approx{\rm 2\,GHz}$ and $Z_F = 115\,\Omega$. Both the CPW and the filter were fabricated by first dc-sputtering 120$\,n$m thick Nb film on top of 500$\,\mu$m thick high resistivity silicon wafer ($\rho > 10\,k \Omega\cdot c$m). This was followed by patterning the design using photolithography and RIE-based dry-etch with \ce{Ar}:\ce{BCl3}:\ce{Cl2}.  Finally, the coupling capacitors $C_{in}$ and $C_{out}$ were made using electron-beam lithography and aluminum evaporation. For this study, the coupling capacitors were made symmetrically (i.e., $C_{in}=C_{out}\equiv C_k$), and three different values were utilized: $C_k\approx 1$, $2$ and $2.5\,f$F, denoted as devices I, II, and III respectively.  The values of $C_k$ were estimated using ANSYS Q3D\cite{ANSYS} for the geometry shown in the inset of Fig. 1b with capacitor overlaps $l = 5, 15, 20\,\mu$m. 

We began the experimental characterization of the CPW-filter design by performing transmission and reflection measurements of the filters fabricated without the CPW (Fig. 2).  These measurements were done in a dip probe at $T = 4.2\,$K using an Agilent N5230A network analyzer.  No cryogenic attenuators or amplifiers were used for these initial measurements. From the transmission measurements, $S_{21}$ is observed to cut off around $2\,{\rm GHz}$, with a roll-off of $60\,{\rm dB/decade}$ up to $4\,{\rm GHz}$, in agreement with both the analytical estimation and Sonnet simulations. For frequencies above $4\,{\rm GHz}$ the simple lumped-element model of the individual filter components can no longer be used to accurately describe the filter's response. In particular, the parasitic capacitance $C_P$ of each inductor is seen to limit the filter's insertion loss: heuristically,  $C_P$  contributes a reactance ($\sim{-1}/{\omega C_P}$) that becomes comparable to the inductive reactance ($\omega L_f$); this causes the total inductor impedance to decrease and ultimately vanish at the self-resonance frequency ($\sim{1}/{\sqrt{L_FC_P}}\simeq12\,{\rm GHz}$), resulting in a maximum in $S_{21}$. More quantitatively, Sonnet\cite{sonnet} simulations predict a maximum in $S_{21}$ at $12\,{\rm GHz}$; the simulated charge and current profiles confirm this maximum corresponds to the lowest-order self-resonance of the inductors - additional simulations show that, with simple design changes to reduce $C_P$, the stop-band could be extended to at least $14\,{\rm GHz}$. Despite the influence of $C_P$, the insertion losses remain in excess of $20\,{\rm dB}$ up to $\sim10\,{\rm GHz}$. Moreover, it is important to note the reflective nature of the filter between $3\,{\rm GHz}$ and $10\,{\rm GHz}$, which is evident from the near unity value of $S_{11}$ measured within the band. 

\begin{figure}
\includegraphics[ width=.9\columnwidth,keepaspectratio]{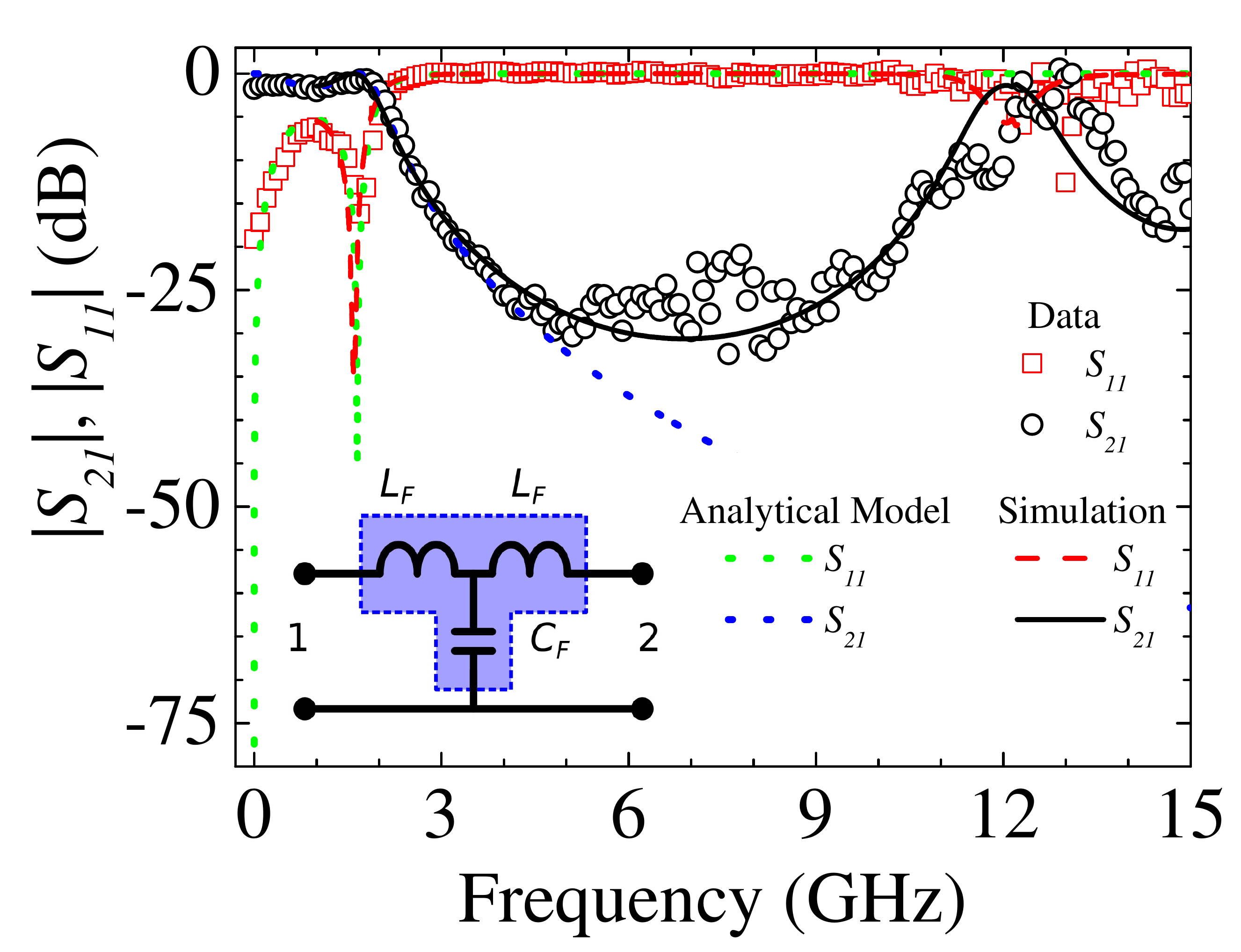}\caption{\footnotesize Measured and simulated transmission ($S_{21}$) and reflection ($S_{11}$) characteristics of a reflective T-filter (see inset) from 0.1 to 15 GHz. The black circles (red squares) represent the magnitude of $S_{21}$ ($S_{11}$) measurements at $4\,{\rm K}$ using a network analyzer calibrated to account for the probe's cable losses and delay.  The solid black (dashed red) line is the magnitude of $S_{21}$ ($S_{11}$) simulated using Sonnet for this filter's geometry.  The dotted lines represent $S_{21}$ and $S_{11}$ calculated with an analytical expression that treats the filter's components as lumped-elements.}\label{fig2}\end{figure}

Next, transmission measurements of complete filtered-CPW circuits were conducted at milli-Kelvin temperatures (Fig. 3). Each sample was mounted in a light-tight copper box and magnetic shield, both of which were anchored to the mixing chamber of a dilution refrigerator; all measurements were made at the base temperature of the fridge, which was measured to be between 20 $m$K and 30 $m$K. An optically-isolated voltage source connected to Port 3 through a heavily-filtered ($>100\,{\rm dB}$ of attenuation at $5\,{\rm GHz}$) coaxial line was used to supply dc voltages, $V_{dc}$, to the devices. The input microwave line to the CPW included $70\,{\rm dB}$ of cryogenic attenuation, which was distributed between refrigerator stages. On the return line, two circulators provided a total of $\sim 30\,{\rm dB}$ of isolation between the CPW and the first-stage amplifier, a cryogenic HEMT, which was anchored to the $4\,{\rm K}$ stage. The output from the HEMT was amplified and mixed down to MHz frequencies using a standard heterodyne circuit, and then digitized using an Alazar ATS9870 card. From the digitized data, the complex transmission $S_{21}$ through the device was recovered and corrected for amplifier gain, cable loss and phase delay. 

\begin{figure}
\begin{center}\includegraphics[ width=.9\columnwidth,keepaspectratio]{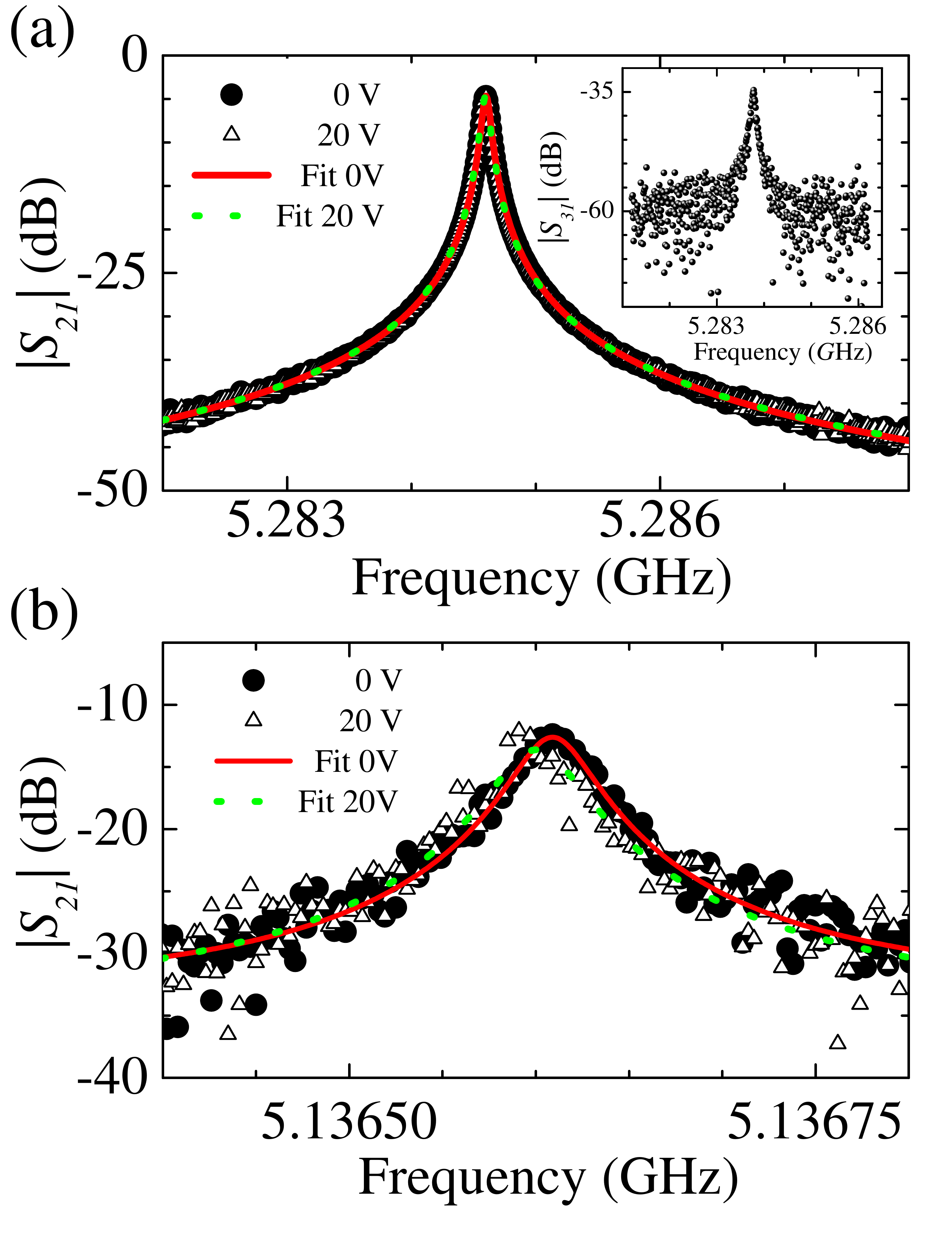}\end{center}\caption{ \footnotesize{\bf a)} Measured $S_{21}$ for a filtered CPW cavity (device II) for $V_{dc}=0 V$ (black circles) and $V_{dc}=20 V$ (open triangles) using high power ($N\approx 10^5$  mean cavity photons).  The data are fit (solid and dotted lines) using the standard expression for a CPW cavity's $S_{21}$  without a filter. The fit yields a loaded quality factor of $Q_{L}=7.4 \times10^{4}$, which we believe to be limited by losses through $C_{in}$ and $C_{out}$ (see Fig. 4). {\bf(inset)} Measurements of $S_{31}$  for Port 3. We note the slight shift in the peak position ($\Delta F = 0.81\,{\rm MHz}$) between the $S_{31}$ and $S_{21}$ measurements, which we believe is a result of the thermal cycling and re-wiring of the device between measurements.   Also, note the difference in scale of the y-axes for $S_{21}$ and $S_{31}$. {\bf b)}  Measured $S_{21}$ for device I in the single-photon regime ($N\simeq1$). From fits to the data, we extract $Q_{L}\simeq1.1 \times10^{5}$, which we believe is limited by loss due to two-level systems (see Fig. 4). For both devices, no resolvable change in $Q_{L}$ is observed over the range of $V_{dc}={\rm 0\,to\,20\,V}$.}\label{fig3}
\end{figure}

Figure 3 (a) shows the amplitude of a transmission spectrum ($S_{21}$) measured for device II at large microwave power, $P=-100\,{\rm dBm}$ at Port 1, corresponding to a mean cavity photon number $N\simeq 10^5$, and for $V_{dc} ={\rm 0\,V}$ and 20V applied to Port 3; Fig. 3(b) displays the corresponding data for low $P$ for Device I ($P=-155\,{\rm dBm}$ and $N\simeq 1$).  The traces are representative of transmission measurements of all the samples studied here, and are well-fit to the standard complex expression\cite{Goppl2008} for $S_{21}$ of a CPW's fundamental mode without the dc filter (solid and dotted lines in the plot).  It is evident from the data that for biases as high as $V_{dc}=20\,$V there is no resolvable change in the shape of the resonance or the mode's quality factor. Also shown are measurements of $S_{31}$ (Fig. 3(a), inset), which is defined as the transmitted signal measured at Port 3 for voltages incident on Port 1  with Port 2 terminated in $50\,\Omega$. The peak of $S_{31}$ is seen to be $\sim 30\,{\rm dB}$ smaller than the corresponding measurements of $S_{21}$, indicating a high degree of isolation through port 3. 

\begin{figure}
\begin{center}\includegraphics[ width=.9\columnwidth,keepaspectratio]{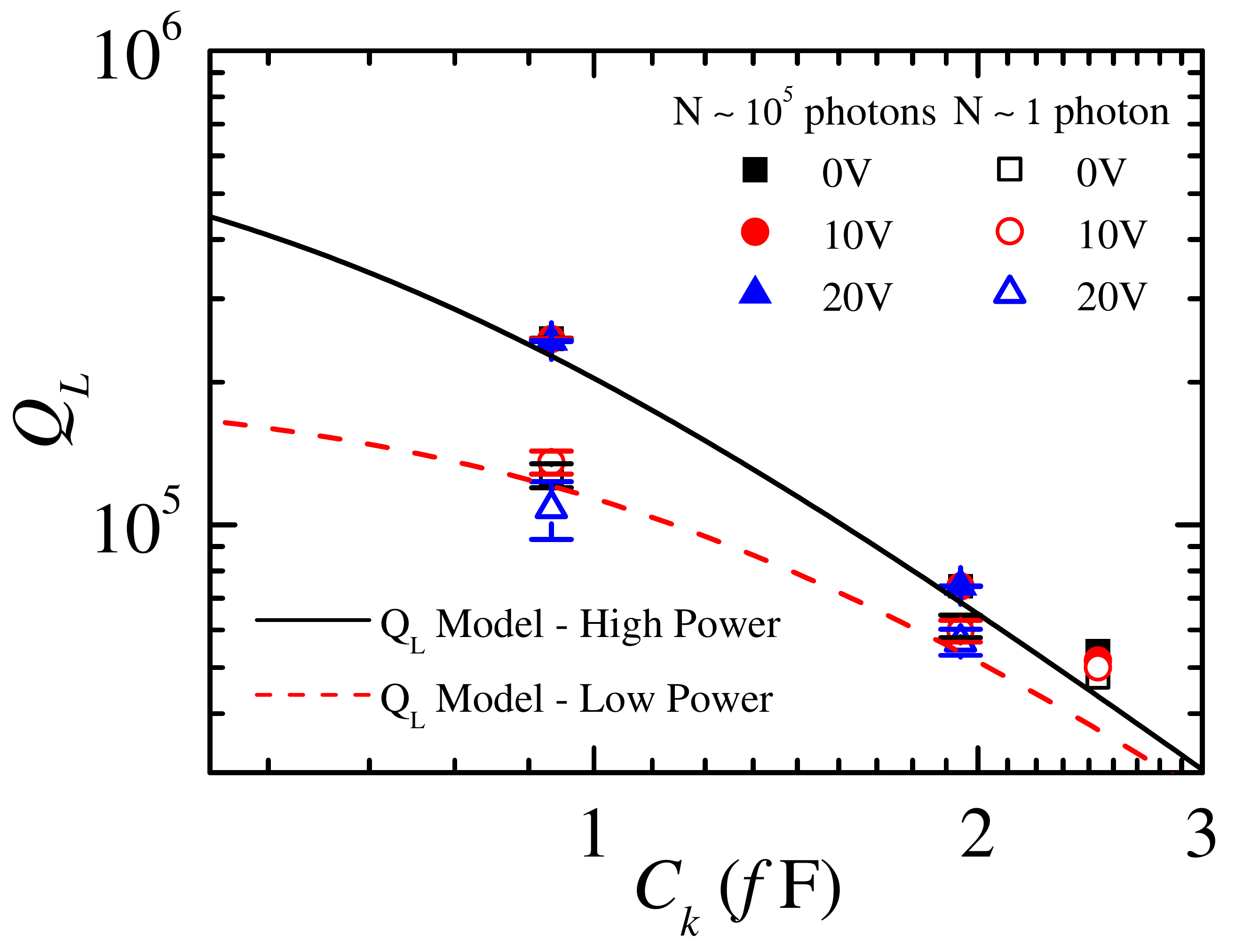}\end{center}\caption{\footnotesize{\bf} Loaded quality factors $Q_L$ of all the devices studied in this work plotted versus $C_{k}$ for the full range of $V_{dc}$ and $P$.  Quality factors were determined from fits to cavity's $S_{21}$ response (e.g. Fig. 3).  Error bars are determined by the $2\sigma$ uncertainty estimated from the fits.}\label{fig4}
\end{figure}

Similar fits of $S_{21}$ data enable extraction of $Q_L$ for the full range of parameters $V_{dc}$, $C_k$, and $P$ that we investigated (Fig. 4).  We find for both high and low $P$ that $Q_L$ increases proportionally to $\sim1/{C_k}^2$, with values as high as  $Q_{L}=2.5 \times10^{5}$ measured for device I. The observations are consistent with estimates of $Q_L$ (solid and dashed lines, Fig. 4), which we define through $1/Q_L \equiv 1/Q_i + 1/Q_C$.  Here $1/Q_C$ accounts for losses to external circuitry through the coupling capacitors.\cite{Goppl2008} Internal cavity losses are represented by ${1}/{Q_{i}}$, which we estimate from the cavities' insertion losses and fits to $S_{21}$; we use $Q_i=0.7\times10^6$ at high power (solid line) and $Q_i=0.2\times10^6$ at low power (dashed line), where unsaturated two-level systems make a greater contribution to the internal dissipation.\cite{Pappas2011} We can estimate the contribution to $Q_{i}$  from the losses through the filter using the ratio  $IL_F={S_{31}(\omega_c)}/{S_{21}(\omega_c)}$. Specifically, for device II (the only device for which $S_{31}$ was measured; Fig. 3(a), inset), we see that $IL_F=-30\,{\rm dB}$, indicating that $0.1\%$ of the power escapes through the filter in comparison with the coupling capacitors. This corresponds to a filter quality factor of $\sim \sqrt{1000} Q_C = 3 \times 10^6$, and suggests that this bias circuit design could be integrated, with small or negligible impact, into cavities with $Q_i$ as high as $\sim 10^6$.

In conclusion, we have reported a simple, effective, and well-modeled filter design for introducing dc and low-frequency voltages into high-quality superconducting CPW cavities.  The high isolation, large stop-band and small cavity loading make this a versatile design that could be integrated  in superconducting CPW cavities and transmission line feeds, or into traces leading to stand-alone superconducting devices for a large range of cryogenic applications.

\vspace{5 mm}

The authors thank B. Plourde for many helpful conversations, advice, and technical assistance. This work was performed in part at the Cornell NanoScale Facility, a member of the National Nanotechnology Infrastructure Network, which is supported by the National Science Foundation (Grant ECCS-0335765). The authors acknowledge support for this work provided by the National Science Foundation under Grant DMR-1056423 and Grant DMR-1312421

%\nocite{*}
\bibliography{Yu_Hao_et_al_CPW_2014}% Produces the bibliography via BibTeX.

\end{document}